\documentclass[aip,jcp,reprint,superscriptaddress,amsmath,amssymb,floatfix]{revtex4-2}

\usepackage{bm,graphicx,booktabs}
\usepackage{amsthm}
\usepackage{amsmath}
\usepackage[colorlinks=true,linkcolor=blue,citecolor=blue,urlcolor=blue]{hyperref}
\graphicspath{{figures/}}

\newtheorem{theorem}{Theorem}
\newtheorem{proposition}[theorem]{Proposition}

\theoremstyle{remark}
\newtheorem{remark}[theorem]{Remark}

\let\vr\relax\newcommand{\vr}{\mathbf r}
\newcommand{\vk}{\mathbf k}
\newcommand{\vn}{\mathbf n}
\newcommand{\vM}{\mathbf M}
\newcommand{\erfc}{\operatorname{erfc}}
\newcommand{\erf}{\operatorname{erf}}
\newcommand{\pLESlong}{projected latent Ewald summation}
\newcommand{\pLES}{pLES}

\begin{document}

\title{Exact periodic electrostatics for latent-charge machine-learning potentials}

\author{Eslam Ibrahim}
\email{eslam.ibrahim@kaust.edu.sa}
\affiliation{Physical Science and Engineering Division (PSE), King Abdullah University of Science and Technology (KAUST), Thuwal, Saudi Arabia}

\begin{abstract}
Latent-charge machine-learning interatomic potentials of the latent Ewald summation (LES) type learn
atomic charges from energies and forces and evaluate their electrostatic energy in reciprocal space.
I formulate this long-range term as a smeared Epstein--Hurwitz lattice sum and derive its exact periodic
structure: a decomposition that isolates a net-charge artefact and a local self-energy, a sphere-average
(image) theorem for cells of any shape, and the forces and virial. Full charge equilibration, LES with
induced charges, constrained and unconstrained LES models and fixed-charge force fields follow as
successive approximations of one variational problem, whose diagonal-hardness limit is a projection of
the latent charges onto fixed total charge; I call the resulting scheme \pLESlong{} (\pLES{}). Energies
and forces determine the latent charges only up to a gauge, so that fragment-level rather than
atom-level charge quantization is the consistent target. The long-wavelength response shows that the
latent channel is guaranteed to be a dielectric only if charge is conserved locally; latent charges that
are local functions of the environment and conserved at most globally can screen like a conductor.
\pLES{} is implemented in the atomic cluster expansion and in LAMMPS, and the exact identities hold to
machine precision, also for the charges of trained models. For charged NaCl clusters, \pLES{} recovers
the total charge that an unconstrained model misses by up to 1.8~$e$ and reduces the test errors of a
short-range model from 1.44 to 0.42~meV/atom and from 37 to 13~meV/\AA{}. In two water datasets,
independently trained constrained and unconstrained models learn the same latent charges up to a global
sign and scale, and the projection keeps every cell exactly neutral.
In molecular dynamics of \pLES{}
water, the latent molecular charges fluctuate and dominate the screening at all accessible wavelengths,
unlike in a fixed-charge water model. Within a crossover model their wavevector dependence excludes a
conductor and gives a latent dielectric constant between 19 and 71, set by compensation of molecular
charges over about 1~nm, a scale that the training cells do not probe. A finite dielectric constant can be
guaranteed only through local charge conservation in the charge model.
\end{abstract}

\maketitle

\section{Introduction}\label{sec:intro}
Machine-learning interatomic potentials (MLIPs) reach near-\emph{ab initio} accuracy at a small fraction
of the cost by writing the energy as a sum of contributions from local atomic
environments~\cite{keith2021,unke2021ffreview,drautz2019,lysogorskiy2021,batzner2022,batatia2022}.
Locality is also their main limitation for ionic, polar and interfacial systems, where Coulomb
interactions decay as $1/r$ and dielectric screening is a collective property of the whole system;
short-range models can then miss long-range structure and response even when their energies and forces
are accurate~\cite{yue2021,kimcheng2026}. Many strategies have been proposed to restore long-range
physics~\cite{baldwin2026,grasselli2026,kimcheng2026,behler2026roadmap}: global or reciprocal-space
representations and message passing~\cite{grisafi2019,huguenin2023,kosmala2023,faller2024}, learned
Wannier centres~\cite{zhang2022dplr}, partial charges fitted to reference charges or
dipoles~\cite{morawietz2012,ko2021,unke2019physnet,unke2021spookynet,anstine2025}, charge equilibration
with learned electronegativities and hardnesses~\cite{rappe1991,ghasemi2015,ko2021,rinaldi2025}, and
self-consistent field models~\cite{gao2022}.

Latent Ewald summation (LES)~\cite{cheng2025les,king2025,zhong2025,kim2025universal} takes a minimal route.
Environment-dependent
latent charges are predicted from local descriptors, smeared as Gaussians and summed with a
reciprocal-space Ewald kernel, and they are learned from energies and forces alone, without
electronic-structure labels. Despite this minimal supervision, the latent charges reproduce Born
effective charges, infrared spectra and the response to external fields~\cite{zhong2025}; the scheme
augments a wide range of short-range architectures~\cite{kim2025universal} and extends to latent
multipoles and non-self-consistent polarization~\cite{kim2026multipoles}.

What has not been formulated is the exact periodic structure of the latent-charge energy and what
energies and forces actually determine about the charges. For classical force fields with fixed charges,
the finite-size electrostatics of periodic cells is well understood~\cite{figueirido1995,hummer1996,%
hunenberger1999,rocklin2013,freysoldt2009}. In a latent-charge model the charges are emergent: their
overall scale is set jointly with the local energy, their total is not conserved unless it is
constrained, and the dielectric response of the long-range channel is a property of the trained model
rather than an input. It is therefore not obvious which part of the latent-charge energy is a cell
artefact, how latent-charge models relate to charge equilibration, which properties of the charges are
physical, how they relate to the topological quantization of charge transport in
insulators~\cite{grasselli2019,pegolo2020}, and whether the latent channel screens like a dielectric at
all.

In this work I write the long-range term of LES-type models as a smeared Epstein--Hurwitz lattice
sum~\cite{nijboer1957}. Section~\ref{sec:model} gives an exact decomposition that isolates the net-charge
artefact and the Gaussian self-energy, together with the forces and virial. Section~\ref{sec:qeq} places
full charge equilibration, LES with induced charges, the constrained and unconstrained LES models and
fixed-charge force fields in one variational hierarchy, in which a projection of the latent charges onto
fixed total charge is the diagonal-hardness limit. Section~\ref{sec:image} proves a sphere-average
(image) theorem for cells of any shape. Section~\ref{sec:gauge} identifies the gauge freedom that energies
and forces leave in the latent charges and shows that fragment-level, not atom-level, charge quantization
is the consistent target. Section~\ref{sec:eps} defines the dielectric response of the latent channel and
shows that it is guaranteed to be that of a dielectric only if charge is conserved locally. I refer to the
latent-charge model with the projection as \pLESlong{} (\pLES{}). Section~\ref{sec:impl} describes its
implementation in the atomic cluster expansion (ACE)~\cite{drautz2019} and in LAMMPS~\cite{thompson2022}
and verifies the exact identities numerically, for test charges and for the charges of a trained model.
Sections~\ref{sec:nacl} and~\ref{sec:water} apply \pLES{} to charged NaCl clusters with non-local charge
redistribution~\cite{ko2021} and to two water datasets, including molecular dynamics of the
long-wavelength response of the latent channel.

\paragraph*{Conventions.} $C=1/(4\pi\varepsilon_0)=14.3996$~eV\,\AA\,$e^{-2}$ and $\beta=1/k_BT$. The
cell matrix is $\mathbf h$, its volume $V=\det\mathbf h$, the direct lattice
$\Lambda=\{\vn=\mathbf h\mathbf m:\mathbf m\in\mathbb Z^3\}$ and the reciprocal lattice
$\Lambda^*=\{\vk=2\pi\mathbf h^{-T}\mathbf m\}$; $R_{\min}=\min_{\vn\neq0}|\vn|$, and for a cubic cell
$L=V^{1/3}=R_{\min}$. Pair vectors $\vr_{ij}$ are minimum-image vectors.

\section{The latent-charge energy as a smeared Epstein--Hurwitz sum}\label{sec:model}
\subsection{Model}
\begin{equation}
E=E_{\rm loc}(\{\vr\})+E_{\rm LR}(\{\vr\},\{q\}),\qquad q_i=q_\theta(\mathcal N_i),
\end{equation}
\begin{equation}
\begin{aligned}
E_{\rm LR}&=\frac{2\pi C}{V}\sum_{\vk\in\Lambda^*\setminus0}\frac{e^{-\sigma^2k^2/2}}{k^2}\,|S(\vk)|^2
=\frac C2\sum_{i,j}q_iq_j\,\Phi_\sigma(\vr_{ij}),\\
S(\vk)&=\sum_iq_ie^{i\vk\cdot\vr_i},\qquad
\Phi_\sigma(\vr)=\frac{4\pi}{V}\sum_{\vk\neq0}\frac{e^{-\sigma^2k^2/2}}{k^2}e^{i\vk\cdot\vr}.
\end{aligned}
\label{eq:Phi}
\end{equation}
The factor $e^{-\sigma^2k^2/2}$ is the Fourier transform of $\erf(\alpha r)/r$ with
$\alpha=1/(\sqrt2\,\sigma)$; the $i=j$ terms are included and finite. The $k$-cutoff $k_c$ and the width
$\sigma$ define the model and must be identical in training and simulation~\cite{cheng2025les,king2025}.

\subsection{Zeta structure and the missing $k=0$ term}
Following the classical lattice-sum treatment~\cite{nijboer1957}, define the Epstein--Hurwitz zeta
function
\begin{equation*}
Z_\Lambda(s;\vr)=\sum_{\vn\in\Lambda}|\vr+\vn|^{-2s},\qquad \operatorname{Re}s>3/2 .
\end{equation*}
Writing $|\mathbf x|^{-2s}=\Gamma(s)^{-1}\int_0^\infty t^{s-1}e^{-t|\mathbf x|^2}dt$, splitting the
integral at $t=\alpha^2$ and applying Poisson summation,
\begin{equation*}
\sum_\vn e^{-t|\vr+\vn|^2}=\frac{\pi^{3/2}}{Vt^{3/2}}\sum_\vk e^{-k^2/4t}e^{i\vk\cdot\vr},
\end{equation*}
gives for all $s\neq3/2$
\begin{equation}
\begin{aligned}
\Gamma(s)Z_\Lambda(s;\vr)={}&\sum_\vn\frac{\Gamma(s,\alpha^2|\vr+\vn|^2)}{|\vr+\vn|^{2s}}
+\frac{\pi^{3/2}}{V}\frac{\alpha^{2s-3}}{s-\frac32}\\
&+\frac{\pi^{3/2}}{V}\sum_{\vk\neq0}\Big(\frac k2\Big)^{2s-3}
\Gamma\!\Big(\tfrac32-s,\tfrac{k^2}{4\alpha^2}\Big)e^{i\vk\cdot\vr}
\end{aligned}
\label{eq:EH}
\end{equation}
independently of $\alpha$. At $s=\frac12$ this is the Ewald potential,
\begin{equation}
\begin{aligned}
\varphi_\Lambda(\vr)\equiv Z_\Lambda(\tfrac12;\vr)={}&\sum_\vn\frac{\erfc(\alpha|\vr+\vn|)}{|\vr+\vn|}
-\frac{\pi}{V\alpha^2}\\
&+\frac{4\pi}V\sum_{\vk\neq0}\frac{e^{-k^2/4\alpha^2}}{k^2}e^{i\vk\cdot\vr},
\end{aligned}
\end{equation}
where $-\pi/V\alpha^2$ is the analytic continuation of the $k=0$ pole. Hence $\int_{\rm cell}\varphi_\Lambda=0$
and $\nabla^2\varphi_\Lambda=-4\pi\sum_\vn\delta(\vr-\vn)+4\pi/V$: zeta regularization is identical to a
uniform neutralizing background. The same formula at $s=3$ gives the $r^{-6}$ lattice sum.

\paragraph*{Self (Wigner) constant.}
\begin{equation}
\begin{aligned}
\psi_\Lambda\equiv{}&\lim_{r\to0}\Big[\varphi_\Lambda(\vr)-\frac1r\Big]\\
={}&\sum_{\vn\neq0}\frac{\erfc(\alpha n)}{n}+\frac{4\pi}V\sum_{\vk\neq0}\frac{e^{-k^2/4\alpha^2}}{k^2}
-\frac{\pi}{V\alpha^2}-\frac{2\alpha}{\sqrt\pi}.
\end{aligned}
\end{equation}
For a simple cubic cell $\psi_\Lambda=\xi_{\rm sc}/L$ with $\xi_{\rm sc}=-2.837297$. For a triclinic cell
$\psi_\Lambda$ is an Epstein zeta value that depends on the cell shape and is cheap to evaluate from
Eq.~\eqref{eq:EH}.

\subsection{Exact decomposition and net latent charge}
With $\alpha=1/(\sqrt2\sigma)$,
\begin{equation}
\Phi_\sigma(\vr)=\varphi_\Lambda(\vr)-\sum_\vn\frac{\erfc(\alpha|\vr+\vn|)}{|\vr+\vn|}+\frac{2\pi\sigma^2}{V}.
\label{eq:decomp}
\end{equation}
Write $\varphi_\Lambda(\vr)-1/r=\psi_\Lambda+2\pi r^2/(3V)+h(\vr)$, where $h$ is harmonic with
$\langle h\rangle_\Omega=0$ for $|\vr|<R_{\min}$ (Sec.~\ref{sec:image}). If all $|\vr_{ij}|<R_{\min}$, up to
image terms of order $e^{-\alpha^2(R_{\min}-r_{ij})^2}$,
\begin{equation}
\begin{aligned}E_{\rm LR}={}&\frac C2\Big(\psi_\Lambda+\frac{2\pi\sigma^2}{V}\Big)Q^2
+\frac{C}{\sqrt{2\pi}\,\sigma}\sum_iq_i^2\\
&+\frac C2\sum_{i\neq j}q_iq_j\Big[\frac{\erf(\alpha r_{ij})}{r_{ij}}+h(\vr_{ij})\Big]
+\frac{\pi C}{3V}\sum_{i\neq j}q_iq_j\,r_{ij}^2\end{aligned}
\label{eq:ELRdecomp}
\end{equation}
with $Q=\sum_iq_i$. The condition holds for any configuration in a cubic cell (minimum-image distances are
at most $\sqrt3L/2$) but can fail in strongly anisotropic cells such as slabs. The first term is a cell
artefact when $Q\neq0$; if $Q$ fluctuates it produces spurious forces
$-C(\psi_\Lambda+2\pi\sigma^2/V)Q\nabla_iQ$.
The second is the Gaussian self-energy; it depends only on each
$q_i$ and is therefore local and absorbable by $E_{\rm loc}$, so keeping or removing it, changes only the local part of the energy.
For a compact cluster, in which every minimum-image vector equals the difference of the chosen positions,
the last term equals $\frac{2\pi C}{3V}\big(Q\sum_iq_ir_i^2-|\vM|^2\big)$ with $\vM=\sum_iq_i\vr_i$, and
contains the tin-foil dipolar coupling $-(2\pi C/3V)|\vM|^2$~\cite{deleeuw1980}. For a bulk liquid that
fills the cell this rewriting does not hold, and the dielectric response follows from the $k\to0$
coupling instead (Sec.~\ref{sec:eps}). For an isolated system, $V\to\infty$ removes $\psi_\Lambda$, the
$\sigma^2/V$ and $r^2$ terms and $h$ (since $c_\ell\sim L^{-(\ell+1)}$, Sec.~\ref{sec:image}), and
Eq.~\eqref{eq:ELRdecomp} reduces to
\begin{equation}
E_{\rm LR}^{\rm open}=\frac C2\sum_{i\neq j}q_iq_j\,\frac{\erf(\alpha r_{ij})}{r_{ij}}
+\frac{C}{\sqrt{2\pi}\sigma}\sum_iq_i^2,
\label{eq:open}
\end{equation}
the real-space Gaussian-charge sum used by LES for molecules~\cite{kim2025universal}.

\paragraph*{Neutrality projection.} Raw outputs $\tilde q_i$ with positive weights $w_i$ are mapped to
\begin{equation}
\begin{aligned}
q_i&=\tilde q_i-\frac{w_i}{W}\Big(\sum_j\tilde q_j-Q_0\Big),\qquad W=\sum_jw_j,\\
\frac{\partial q_i}{\partial\tilde q_j}&=\delta_{ij}-\frac{w_i}{W}\quad (w\ \text{fixed}),
\end{aligned}
\label{eq:projection}
\end{equation}
which fixes $Q=Q_0$. In neutral cells ($Q_0=0$) this removes the artefact and the spurious forces above.
In charged cells it removes the spurious forces, while the configuration-independent term
$\frac C2(\psi_\Lambda+2\pi\sigma^2/V)Q_0^2$ remains; it depends on the volume and shape of the cell,
contributes to the virial, and must be corrected explicitly in variable-cell simulations of charged cells.
If the weights depend on the atomic environments, their derivatives add terms to Eqs.~\eqref{eq:forces}
and~\eqref{eq:virial} below; in the present implementation they are species constants. Uniform corrections of the
total charge are also used in PhysNet~\cite{unke2019physnet} and qNEP~\cite{fan2026qnep}. 
Here the
projection follows from the variational hierarchy of Sec.~\ref{sec:qeq}, which identifies species-dependent
weights as inverse hardnesses and the projection as charge equilibration with a diagonal hardness matrix.
The value of $Q_0$ for a charged cell is set by the gauge condition of Sec.~\ref{sec:gauge}.

\subsection{Forces and virial}\label{sec:forces}
  With
  \begin{equation*}
  \phi_j=\frac{\partial E_{\rm LR}}{\partial q_j}
  =\frac{4\pi C}{V}\sum_{\vk\neq0}\frac{e^{-\sigma^2k^2/2}}{k^2}\operatorname{Re}\big[S^*(\vk)e^{i\vk\cdot\vr_j}\big],
  \end{equation*}
\begin{equation}
\begin{aligned}
\mathbf F_i={}&\frac{4\pi C}{V}q_i\sum_{\vk\neq0}\frac{e^{-\sigma^2k^2/2}}{k^2}\,\vk\,
\operatorname{Im}\big[S^*(\vk)e^{i\vk\cdot\vr_i}\big]\\
&-\sum_j\phi_j\frac{\partial q_j}{\partial\vr_i}-\frac{\partial E_{\rm loc}}{\partial\vr_i}.
\end{aligned}
\label{eq:forces}
\end{equation}
Under homogeneous strain $\epsilon$, $\vk\to(1-\epsilon^T)\vk$ with $\vk\cdot\vr$ invariant. With $E_\vk$
the summand of $E_{\rm LR}$,
\begin{equation}
\begin{aligned}
W_{\alpha\beta}=-\frac{\partial E_{\rm LR}}{\partial\epsilon_{\alpha\beta}}={}&\sum_{\vk\neq0}E_\vk
\Big[\delta_{\alpha\beta}-2k_\alpha k_\beta\Big(\frac1{k^2}+\frac{\sigma^2}2\Big)\Big]\\
&-\sum_j\phi_j\frac{\partial q_j}{\partial\epsilon_{\alpha\beta}},
\end{aligned}
\label{eq:virial}
\end{equation}
which is required for stress training and for $NPT$ simulations.

\section{Relation to charge equilibration}\label{sec:qeq}
LES-type models and charge-equilibration (QEq) models~\cite{rappe1991,ghasemi2015,ko2021,rinaldi2025}
both assign environment-dependent atomic charges, by very different procedures. Their charges are members
of one hierarchy, obtained from a single variational problem by successive approximations. Consider the
generalized QEq energy and the corresponding Lagrangian
\begin{equation}
\begin{aligned}
E_{\rm QEq}[q]&=\sum_i\Big(\chi_iq_i+\tfrac12\eta_iq_i^2\Big)+\frac12\sum_{i\neq j}q_iJ_{ij}q_j,\\
\Omega[q]&=E_{\rm QEq}[q]-\mu\Big(\sum_iq_i-Q_0\Big),
\end{aligned}
\label{eq:qeq}
\end{equation}
with environment-dependent electronegativities $\chi_i$ and hardnesses $\eta_i>0$ (local functions of
$\mathcal N_i$, with the Gaussian self-energy $C\Phi_\sigma(0)$ absorbed into $\eta_i$), the periodic
smeared Coulomb kernel $J_{ij}=C\Phi_\sigma(\vr_{ij})$ for $i\neq j$, and a Lagrange multiplier $\mu$, the
electron chemical potential. With $D={\rm diag}(\eta_i)$ and $H=D+J$, stationarity of $\Omega$ gives
\begin{equation}
Hq=\mu\mathbf 1-\boldsymbol\chi,\qquad \sum_iq_i=Q_0 .
\label{eq:qeqstat}
\end{equation}
The stationary point is a unique minimum if $H$ is positive definite on the subspace of fixed total charge;
positive hardnesses alone do not ensure this, because the Coulomb coupling can make $H$ indefinite.

\paragraph*{(i) Full charge equilibration.} $q=H^{-1}(\mu\mathbf 1-\boldsymbol\chi)$ with
$\mu=(Q_0+\mathbf 1^TH^{-1}\boldsymbol\chi)/(\mathbf 1^TH^{-1}\mathbf 1)$. The charges are non-local,
charge can be transferred across the whole cell, and a linear system must be solved at every step, as in
4G-HDNNP and charge-constrained ACE~\cite{ko2021,rinaldi2025}. If the energy contains $E_{\rm QEq}$ at its
constrained stationary point, the charge derivatives drop out of the forces:
$\sum_j(\partial E_{\rm QEq}/\partial q_j)(\partial q_j/\partial\vr_i)=\mu\sum_j\partial q_j/\partial\vr_i=0$
at fixed $Q_0$.

\paragraph*{(ii) First-order expansion: induced charges.} Expanding $H^{-1}=D^{-1}-D^{-1}JD^{-1}+O(J^2)$
at fixed $\mu$ gives $q=q^{(0)}+\Delta q$ with
\begin{equation}
\begin{aligned}
q_i^{(0)}&=\frac{\mu-\chi_i}{\eta_i},\qquad
\Delta q_i=-\frac1{\eta_i}\sum_{j}J_{ij}q^{(0)}_j=-\kappa_i\,\Phi(\vr_i),\\
&\kappa_i=\eta_i^{-1},
\end{aligned}
\end{equation}
where $\Phi(\vr_i)$ is the potential of the zeroth-order charges at atom $i$: a non-self-consistent charge
induced in proportion to the local potential. This is the form of the induced-charge term of the
polarizable-multipole extension of LES~\cite{kim2026multipoles}, where it is used without a charge
constraint, with a separately learned hardness and the potential of fixed latent multipoles, and where it
is noted to metallize insulators, as charge equilibration does.
At fixed $Q_0$ instead, the first-order charges are obtained, to first order in $J$, by applying the
projection Eq.~\eqref{eq:projection} with $w_i=1/\eta_i$ to $q^{(0)}+\Delta q$. The series converges when
the spectral radius of $D^{-1}J$ is below one and diverges in the conducting limit (metals, electrodes).

\paragraph*{(iii) Diagonal limit: the neutrality projection.} Omitting $J$ from the charge-determining
step gives $q_i=(\mu-\chi_i)/\eta_i$ with $\sum_iq_i=Q_0$, which is exactly the projection
Eq.~\eqref{eq:projection} with
\begin{equation}
\tilde q_i=-\chi_i/\eta_i,\qquad w_i=1/\eta_i .
\end{equation}
The charges depend on local environments and on a single global multiplier; the cost is one global
reduction, with no linear solve. The Coulomb energy $E_{\rm LR}$ is still evaluated with these charges.
Because they are not stationary points of $E_{\rm LR}$, the forces contain
$-\sum_j\phi_j\,\partial q_j/\partial\vr_i$ (Sec.~\ref{sec:forces}); this single back-propagation is the
price for avoiding the solve.

\paragraph*{(iv) Fixed chemical potential.} Fixing $\mu$ instead of $Q_0$ gives $q_i=\tilde q_i+w_i\mu$.
The case $\mu=0$ is the original, unconstrained LES model~\cite{cheng2025les,king2025}.

\paragraph*{(v) Fixed charges.} Environment-independent $\chi_i$ and $\eta_i$ give fixed species
charges, i.e.\ classical force fields. With charges proportional to oxidation numbers this is the limit
enforced by atom-level quantization (Proposition~\ref{prop:collapse}).

\begin{proposition}[Charge-equilibration hierarchy]\label{prop:hierarchy}
The charges of full charge equilibration (i), its first-order expansion with induced charges (ii), the
constrained LES model with neutrality projection (iii), the unconstrained LES model (iv, $\mu=0$) and
fixed-charge models (v) are successive approximations of the stationary point of Eq.~\eqref{eq:qeq},
obtained by truncating the Coulomb coupling $J$ in the charge-determining step and by fixing either $Q_0$
or $\mu$.
\end{proposition}

The hierarchy concerns the charges, not the energy. Evaluated at the charges of level (iii), the on-site
QEq energy is
$\sum_i(\chi_iq_i+\frac12\eta_iq_i^2)=-\frac12\sum_i\eta_i\tilde q_i^2+(\sum_i\tilde q_i-Q_0)^2/2W$; the first term
is local and absorbable by $E_{\rm loc}$, whereas the second is a non-local penalty on the mismatch
$\sum_i\tilde q_i-Q_0$ that the \pLES{} energy $E_{\rm loc}+E_{\rm LR}$ does not contain.

\paragraph*{What the diagonal limit gives up and what it keeps.} The diagonal limit has no charge transfer
driven by the electrostatic potential and therefore no field-driven charge-flow polarizability, the
origin of the superlinear size scaling of the polarizability in QEq~\cite{warren2008,jensen2023};
electronic screening then enters only through the scale of the latent charges (Sec.~\ref{sec:gauge}). It
shares with QEq a single chemical potential for the whole system: any mismatch $\sum_j\tilde q_j-Q_0$ is
distributed over all atoms in proportion to $w_i$. Charge is therefore conserved globally but not
locally. A fragment whose environment changes acquires a latent charge from its local structure, and the
opposite charge is spread over the whole cell. The total charge of a well-separated fragment is therefore
correct only if the model has learned it fragment by fragment (Sec.~\ref{sec:gauge}), and the dielectric
response of the latent channel is not guaranteed to be that of a dielectric (Sec.~\ref{sec:eps}). The same
holds at levels (i) and (iv).

\section{Image theorem}\label{sec:image}
\begin{theorem}[Image theorem]\label{thm:image}
Let $g(\vr)=\varphi_\Lambda(\vr)-1/r-2\pi r^2/(3V)$. Then $\nabla^2g=-4\pi\sum_{\vn\neq0}\delta(\vr-\vn)$, so
$g$ is harmonic in $|\vr|<R_{\min}$, and for any cell shape
\begin{equation}
\langle\varphi_\Lambda\rangle_\Omega(r)=\frac1r+\psi_\Lambda+\frac{2\pi r^2}{3V},\qquad r<R_{\min}.
\label{eq:image}
\end{equation}
\end{theorem}
\begin{proof}
$\nabla^2(2\pi r^2/3V)=4\pi/V$ cancels the background term of $\nabla^2\varphi_\Lambda$, and
$\nabla^2(1/r)=-4\pi\delta(\vr)$ removes the $\vn=0$ source. By the mean-value property of harmonic
functions, $\langle g\rangle_\Omega(r)=g(0)=\psi_\Lambda$.
\end{proof}
The cell shape enters only through $\psi_\Lambda$; the $r^2$ coefficient depends only on $V$. The constant
and quadratic terms of the Ewald potential are known from the finite-size analysis of ionic solvation in
cubic cells~\cite{figueirido1995,hummer1996}; the mean-value argument extends the sphere average to
arbitrary cells and makes the remainder explicit. For the smeared kernel,
\begin{equation}
\begin{aligned}
\langle\Phi_\sigma\rangle_\Omega(r)={}&\frac{\erf(r/\sqrt2\sigma)}{r}+\psi_\Lambda\\
&+\frac{2\pi r^2}{3V}+\frac{2\pi\sigma^2}{V}+O\big(e^{-\alpha^2(R_{\min}-r)^2}\big).
\end{aligned}
\end{equation}
\paragraph*{Anisotropic remainder.} $h=\sum_{\ell\ge2}c_{\ell m}r^\ell Y_{\ell m}$ with
$c_\ell\sim L^{-(\ell+1)}$; odd $\ell$ vanish by inversion symmetry, cubic cells start at $\ell=4$ and
non-cubic cells at $\ell=2$. Its first-order contribution to an orientationally averaged quantity
vanishes when all orientations of $\vr_{ij}$ are sampled, which holds for $r$ below the inradius of the
Wigner--Seitz cell ($L/2$ for a cubic cell, possibly small for slabs).

\section{Gauge freedom of latent charges}\label{sec:gauge}
\subsection{What energies and forces determine}
Energies and forces constrain the latent charges only through $E_{\rm LR}+E_{\rm loc}$. This leaves three
kinds of indeterminacy of different nature: an exact symmetry (the sign), a quantity that local models
cannot represent (the net charge), and freedoms that are not exact but only weakly constrained by
finite training data (the scale and the atomic partition).

\paragraph*{Sign.} $E_{\rm LR}$ is quadratic in $q$, so $q\to-q$ leaves it unchanged. For the unconstrained
model ($\mu=0$) this is an exact symmetry, and the overall sign of the learned charges is set by the
initialization. With the projection, $q\to-q$ maps $Q_0\to-Q_0$: the symmetry is broken in charged
systems and remains for neutral ones.

\paragraph*{Scale and partition.} A change of the latent charges that alters $E_{\rm LR}$ only through
short-range terms can be compensated by $E_{\rm loc}$. Self-energies and redistribution between neighbouring
atoms are therefore set jointly with the local energy. A uniform rescaling of all charges is not an exact
symmetry: it changes the interaction between well-separated fragments, which is observable in principle.
Training cells of about 1~nm, however, constrain it only weakly (Sec.~\ref{sec:water}). This agrees with
the finding that LES charges reproduce Born effective charges after a single rescaling by
$\sqrt{\varepsilon_\infty}$~\cite{zhong2025}: the scale is fixed by the dielectric response
(Eq.~\eqref{eq:gauge} and Remark~\ref{rem:scaling}), not by energies and forces of small cells. Atomic latent charges are not
observables, as are not the atomic charges of any partition of the electron density.

\paragraph*{Net charge.} Without the projection, $Q=\sum_i\tilde q_i$ is a sum of local contributions and
cannot represent a total charge that depends on the system as a whole. The projection supplies it
exactly.

\subsection{Gauge condition}
For well-separated fragments $A,B$ with $Q_A^{\rm lat}=\sum_{i\in A}q_i$, the effective interaction of the
model decays as $w_{AB}(r)\to CQ_A^{\rm lat}Q_B^{\rm lat}/(\varepsilon_{\rm lat}r)$, with $\varepsilon_{\rm lat}$
the dielectric constant of the latent channel (Sec.~\ref{sec:eps}), provided that it is finite. Matching
the physical interaction $Cz_Az_B/(\varepsilon r)$ for every pair of species requires
\begin{equation}
Q_A^{\rm lat}=\lambda z_A\ \ \forall A,\qquad\lambda=\sqrt{\varepsilon_{\rm lat}/\varepsilon}
\label{eq:gauge}
\end{equation}
with $z_A$ the integer charge, a single $\lambda$, and $\varepsilon$ the dielectric constant of the reference
electronic-structure method. For a charged periodic cell with physical charge $Z$, the projection then
uses $Q_0=\lambda Z$; for an isolated system in vacuum $\lambda=1$ and $Q_0=Z$, which is the case tested in
Sec.~\ref{sec:nacl}. Testing Eq.~\eqref{eq:gauge} in solution requires ions and the dielectric constant of
the reference method and is left for future work.

\begin{remark}[Relation to charge scaling]\label{rem:scaling}
If the latent channel carries only the nuclear-orientational part of the response,
$\varepsilon_{\rm lat}=\varepsilon/\varepsilon_\infty$ and $\lambda=\varepsilon_\infty^{-1/2}$. This is the
electronic continuum correction of nonpolarizable force fields~\cite{leontyev2011} and the scaling
$q^{\rm les}=q/\sqrt{\varepsilon_\infty}$ of latent charges in the LES
framework~\cite{zhong2025,kim2025universal,kim2026multipoles}. Equation~\eqref{eq:gauge} contains both as
the special case in which the electronic screening is absorbed into the charges. A single scalar $\lambda$
presupposes homogeneous, isotropic screening; for heterogeneous systems such as interfaces a single
screening factor is not generally justified~\cite{kim2026multipoles}.
\end{remark}

\subsection{Atom-level versus fragment-level quantization}
The latent Born effective charges and the transported dipole are
\begin{equation}
\begin{aligned}
Z^*_{i,\alpha\beta}&=\frac{\partial M_\alpha}{\partial r_{i\beta}}=q_i\delta_{\alpha\beta}
+\sum_jr_{j\alpha}\frac{\partial q_j}{\partial r_{i\beta}},\\
\Delta\vM_\gamma&=\oint_\gamma\sum_iZ^*_i\,d\vr_i .
\end{aligned}
\end{equation}
In electronic insulators, a loop that carries atom $j$ across a lattice vector $\vn$ while all other atoms
return transports $\Delta\vM=o_j\vn$, with $o_j\in\mathbb Z$ the oxidation number, provided the gap stays
open along the loop (strong adiabaticity)~\cite{grasselli2019,pegolo2020}. This theorem concerns the
electronic polarization; latent charges need not obey it. The following proposition shows what imposing it
on the latent polarization $\vM=\sum_iq_i\vr_i$ would imply.

\begin{proposition}[Atom-level quantization collapses latent charges]\label{prop:collapse}
Let $\vM=\sum_iq_i\vr_i$ with periodic, environment-dependent $q_i$ and unwrapped $\vr_i$. For the
single-atom loop above, $\Delta\vM=q_j\vn$ exactly. Requiring $\Delta\vM=\lambda o_j\vn$ therefore forces
$q_j=\lambda o_j$ at every configuration that can be a loop endpoint.
\end{proposition}
\begin{proof}
The final configuration differs from the initial one only by the lattice translation of $\vr_j$. All
environments, and therefore all $q_i$, including projected ones, are unchanged, and $\Delta\vM=q_j\vn$.
\end{proof}

\begin{remark}[Atom-level quantization is too strong]\label{rem:contradiction}
Oxidation numbers do not change in acid--base chemistry, so Proposition~\ref{prop:collapse} reduces the
model to fixed charges $\lambda\times$(oxidation number), i.e.\ level (v) of the hierarchy. For water, a
point-charge model of this type reproduces realistic molecular dipoles only with $\lambda$ well below one
(SPC water is $0.41\times(+1,-2)$~\cite{berendsen1981}), whereas Eq.~\eqref{eq:gauge} ties $\lambda$ to the
ratio of dielectric constants. A single $\lambda$ is unlikely to satisfy both.
\end{remark}

\begin{remark}[Fragment-level quantization is the consistent target]\label{rem:fragment}
Atom-level consistency requires auxiliary sites that carry the polarization while all charges remain
integers, as in DPLR~\cite{zhang2022dplr}. For latent point-charge models the consistent target is
fragment-level quantization, $Q_A^{\rm lat}=\lambda z_A$ for every well-separated ion or molecule, with
charge free to redistribute within each fragment. Neutral water is then unconstrained internally, its
molecular charges are conserved, and the latent channel has a finite $\varepsilon_{\rm lat}$ under the
conditions of Proposition~\ref{prop:screening}(a).
Quantized transport along Grotthuss loops is not guaranteed, because $\Delta\vM$ then depends on the latent
charge of the hydrogen that winds across the cell. A fragment-level condition can be added to the loss
function as a penalty; this reduces, but does not remove, the fluctuations of the fragment charges.
\end{remark}

\section{Dielectric response of the latent channel}\label{sec:eps}
\paragraph*{Definition.} The macroscopic response of the model enters only through the $k\to0$ coupling of
$E_{\rm LR}$, $v(k)=4\pi Ce^{-\sigma^2k^2/2}/k^2$. An external potential coupled to the latent charges is a
function of the positions only, also when the charges depend on the environment and are projected, so
classical linear response gives the longitudinal dielectric function of the latent channel
exactly~\cite{hansen2013},
\begin{equation}
\frac1{\varepsilon_L(k)}=1-\chi(k),\qquad
\chi(k)\equiv\frac{\beta v(k)}{V}\langle|S(\vk)|^2\rangle ,
\label{eq:sumrule}
\end{equation}
with $v(k)\to4\pi C/k^2$ at small $k$. A dielectric has $\chi(k)\to1-\varepsilon_{\rm lat}^{-1}$; with free
ions $\chi\to1$ (Stillinger--Lovett perfect screening~\cite{stillinger1968}). In a cubic cell
$k_m^2=(2\pi/L)^2m$ with $r_3(m)$ degenerate vectors ($r_3$ is the sum-of-three-squares function; the
shells $m=4^a(8b+7)$ are empty), so the limit requires more than one box size.

\paragraph*{Screening criterion.} Equivalently, $\varepsilon_L(k)=1+v(k)\bar\chi(k)$, where
$\bar\chi(k)=\beta\langle|S(\vk)|^2\rangle_{\rm p}/V$ is the proper (irreducible) response; at small $k$,
where the collective charge mode is Gaussian, it equals the charge fluctuations without the long-range
part of the coupling~\cite{hansen2013}. Grouping the atoms into fragments $A$ (molecules) with centres
$\vr_A$, charges $Q_A$ and dipoles $\boldsymbol\mu_A$,
\begin{equation}
S(\vk)=\sum_Ae^{i\vk\cdot\vr_A}Q_A+i\vk\cdot\sum_Ae^{i\vk\cdot\vr_A}\boldsymbol\mu_A+O(k^2).
\label{eq:Sfrag}
\end{equation}
Whether the first term survives at long wavelength decides the response.

\begin{proposition}[Local charge conservation and screening]\label{prop:screening}
Consider an ion-free system.
(a) If every fragment is neutral at all times, or charge moves only between fragments within a finite
distance, the charge density is the divergence of a local polarization $\mathbf P$ and $\bar\chi(k)=O(k^2)$.
If, in addition, the long-wavelength proper polarization susceptibility is finite (no ferroelectric or
critical divergence), the latent channel is a dielectric with finite $\varepsilon_{\rm lat}$. Then Eq.~\eqref{eq:sumrule} and
the fluctuations of the total polarization $\vM=\mathbf P(\vk=0)$ with tin-foil
boundaries~\cite{neumann1983},
\begin{equation}
\varepsilon_{\rm lat}=1+\frac{4\pi\beta C}{3V}\big(\langle|\vM|^2\rangle-|\langle\vM\rangle|^2\big),
\label{eq:neumann}
\end{equation}
give the same $\varepsilon_{\rm lat}$.
(b) If the fragment charges are local functions of the environment, charge is conserved at most globally,
and the proper fragment-charge fluctuations have a finite long-wavelength limit, $\bar\chi(0)>0$, then
\begin{equation}
\varepsilon_L(k)=\varepsilon_d+\frac{k_s^2}{k^2}+O(k^2),\qquad
k_s^2=\frac{4\pi\beta C}{V}\lim_{k\to0}\langle|S(\vk)|^2\rangle_{\rm p},
\label{eq:conductor}
\end{equation}
so that $\chi\to1$ and the latent channel screens like a conductor even without ions.
\end{proposition}
\begin{proof}
(a) With conserved or locally transferred charge, the first term of Eq.~\eqref{eq:Sfrag} is $i\vk$ times
the transfer polarization, so $S(\vk)=i\vk\cdot\mathbf P(\vk)+O(k^2)$, and with a finite polarization
susceptibility $v\bar\chi$ has a finite limit, $\varepsilon_{\rm lat}-1$; the equivalence of the longitudinal and tin-foil dipolar routes is the standard
result for a polarization field~\cite{neumann1983,hansen2013}. (b) With $\bar\chi(0)>0$,
$v\bar\chi\to4\pi C\bar\chi(0)/k^2$.
\end{proof}
The assumption of case (b) is generic: without a local conservation law, nothing forces
$\sum_AQ_Ae^{i\vk\cdot\vr_A}$ to vanish as $k\to0$, but a learned charge model can approach case (a)
if its fragment-charge fluctuations happen to be compensated locally. The projection [level (iii)], the
unconstrained LES model [level (iv)] and QEq with one chemical potential [level (i)] lack a local
conservation law. In \pLES{}, a fragment acquires latent charge from its local structure and the
projection spreads the opposite charge over the whole cell; this has no current and no polarization, so
Eq.~\eqref{eq:neumann} has no counterpart. In case (b), two latent test charges interact through
$v(k)/\varepsilon_L(k)\to4\pi C/(\varepsilon_dk^2+k_s^2)$, i.e.\ $C\,e^{-r/\ell}/(\varepsilon_dr)$ with
$\ell=\sqrt{\varepsilon_d}/k_s$, instead of $C/(\varepsilon_{\rm lat}r)$.

\paragraph*{Practical test.} A fit of $\chi(k)=1-\varepsilon_{\rm lat}^{-1}+ak^2$ presupposes that
$\varepsilon_L(k)$ has converged at the sampled wavevectors. The cases are distinguished by
$k^2\varepsilon_L(k)=k^2/(1-\chi)$, which extrapolates to zero for a dielectric and to $k_s^2$ in case (b);
a dielectric whose crossover lies below the smallest accessible wavevector is described by
$1-\chi=\varepsilon_{\rm lat}^{-1}+k^2/(k_s^2+\varepsilon_dk^2)$. Neutralizing each fragment,
$q_i\to q_i-Q_A/n_A$, gives a well-defined dipole, but these are not the charges coupled in $E_{\rm LR}$:
neither Eq.~\eqref{eq:sumrule} nor Eq.~\eqref{eq:neumann} for the neutralized charges then describes the
latent channel, and the two need not agree with each other.

\section{Implementation and numerical verification}\label{sec:impl}
\subsection{\pLES{} in the atomic cluster expansion}
In ACE with Finnis--Sinclair (FS) type embedding~\cite{drautz2019,lysogorskiy2021,finnis1984}, the atomic
energy is $E_i=\sum_{p=1}^{n_E}F_p(\rho_i^{(p)})$, where the densities $\rho_i^{(p)}$ are linear in the ACE
basis functions of the environment of atom $i$. I append $n_q$ further densities of the same basis and
obtain the raw latent charge from the same embedding,
\begin{equation}
\tilde q_i=\sum_{p=n_E+1}^{n_E+n_q}F_p\big(\rho_i^{(p)}\big),
\label{eq:qace}
\end{equation}
so that the charge shares the descriptor of the energy at negligible extra cost. The charges are mapped by
Eq.~\eqref{eq:projection} and enter $E_{\rm LR}$, which is evaluated in reciprocal space for periodic cells,
with $e^{i\vk\cdot\vr}$ factorized into phase factors along the three reciprocal lattice vectors, and with
Eq.~\eqref{eq:open} for isolated systems. Because $E_{\rm LR}$ is quadratic in $q$, $q=0$ is a stationary
point of the loss, and the charge coefficients are initialized randomly with a small scale (standard deviation
$10^{-3}$ for liquid water and $10^{-2}$ for the water phases). Forces and
stresses follow by automatic differentiation in TensorPotential, and the models are fitted with
\textsc{pacemaker}~\cite{bochkarev2022}.

For molecular dynamics the charge-derivative term of Eq.~\eqref{eq:forces} is evaluated without explicit
charge derivatives. With Eq.~\eqref{eq:projection} at fixed weights,
   \begin{equation}
   \begin{aligned}
   \sum_j\phi_j\frac{\partial q_j}{\partial\vr_i}&=\sum_k\phi'_k\frac{\partial\tilde q_k}{\partial\vr_i},\\
   \phi'_k=\phi_k-\bar\phi_w,\qquad \bar\phi_w&=\frac1W\sum_jw_j\phi_j,
   \end{aligned}
   \end{equation}
and by Eq.~\eqref{eq:qace} the right-hand side is the standard ACE back-propagation with the embedding
derivatives $F'_p$ of the charge densities multiplied by $\phi'_k$. The LAMMPS pair style
\texttt{pace/les} therefore performs a charge-only ACE pass, global reductions of $\sum_i\tilde q_i$, $W$ and
$S(\vk)$, the reciprocal-space energy, forces and virial, and a second ACE pass with the weighted charge
derivatives. Parallelization adds only global reductions to the existing ACE domain decomposition, and no
\texttt{kspace\_style} is used. The reciprocal-space sum is evaluated directly; its cost grows as
$N^2$ at fixed density, and a particle--mesh evaluation for large systems is left for future work.

\subsection{Numerical verification}
\begin{figure*}
  \includegraphics[width=\textwidth]{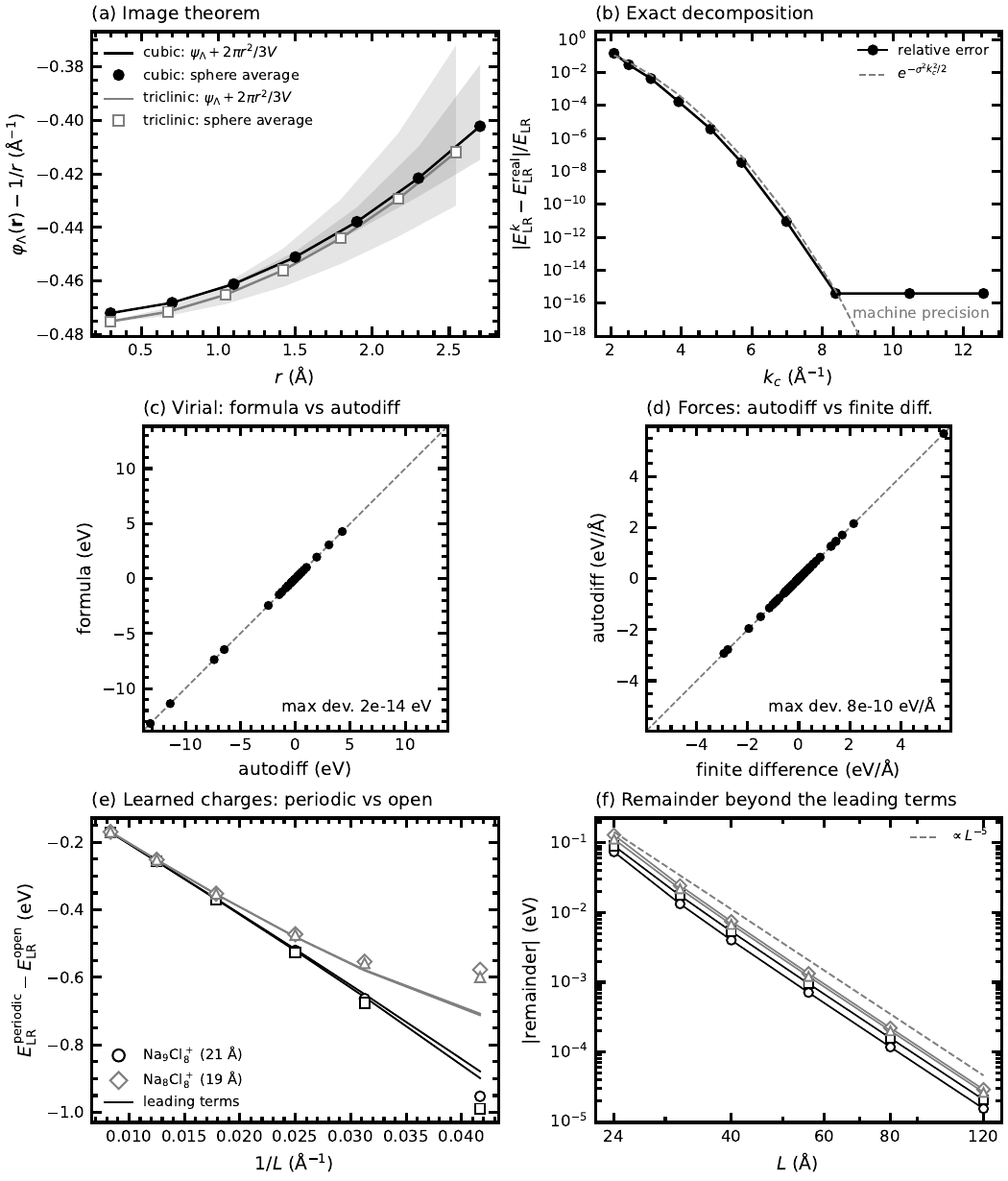}
  \caption{Numerical verification of the exact identities.
  (a) Image theorem: the sphere average of $\varphi_\Lambda(\vr)-1/r$ (symbols) follows
  $\psi_\Lambda+2\pi r^2/3V$ (lines) for a cubic and a triclinic cell; shaded bands show the spread
  over single directions, i.e.\ the anisotropic remainder.
  (b) Relative difference between the reciprocal-space and the real-space evaluation of $E_{\rm LR}$
  as a function of the cutoff $k_c$; it follows the truncation error $e^{-\sigma^2k_c^2/2}$ down to
  machine precision.
  (c) Virial from the analytical formula vs automatic differentiation and (d) forces from automatic
  differentiation vs central finite differences, for random fixed charges.
  (e) $E_{\rm LR}^{\rm periodic}-E_{\rm LR}^{\rm open}$ for the learned \pLES{} charges of two
  Na$_9$Cl$_8^+$ (circles) and two Na$_8$Cl$_8^+$ (diamonds) test clusters in cubic cells of length $L$
  (symbols), and the leading terms $\frac C2(\psi_\Lambda+2\pi\sigma^2/V)Q^2+\frac{\pi C}{3V}\sum_{i\neq j}
  q_iq_jr_{ij}^2$ of Eq.~\eqref{eq:ELRdecomp} (lines).
  (f) Remainder beyond the leading terms; dashed: $\propto L^{-5}$.}
  \label{fig:identities}
\end{figure*}

Figure~\ref{fig:identities} verifies the exact results of Secs.~\ref{sec:model} and~\ref{sec:image}. The
simple-cubic constant $\psi_\Lambda L=-2.8372975$ is reproduced for $L=3$, $6$ and $12$~\AA. The sphere
average of $\varphi_\Lambda(\vr)-1/r$ agrees with $\psi_\Lambda+2\pi r^2/3V$ to $4\times10^{-7}$~\AA$^{-1}$ in a
cubic and a triclinic cell, while individual directions deviate by up to $6\times10^{-2}$~\AA$^{-1}$
[Fig.~\ref{fig:identities}(a)]. The reciprocal-space energy converges to the real-space route of
Eq.~\eqref{eq:decomp} as $e^{-\sigma^2k_c^2/2}$, down to a relative error of $4\times10^{-16}$
[Fig.~\ref{fig:identities}(b)]. For random fixed charges, the virial of Eq.~\eqref{eq:virial} agrees with
automatic differentiation to $2\times10^{-14}$~eV, and the forces agree with central finite differences to
$8\times10^{-10}$~eV/\AA{} [Fig.~\ref{fig:identities}(c,d)]; the charge-derivative terms of
Eqs.~\eqref{eq:forces} and~\eqref{eq:virial} are tested by the comparison with LAMMPS below.

The decomposition also holds for the charges of a trained model. I placed four NaCl test clusters with
their learned \pLES{} charges (Sec.~\ref{sec:nacl}; $Q=+1$, 19--21~\AA{} across) in cubic cells of
$L=24$--$120$~\AA, so that all pair distances are below $L$ and Eq.~\eqref{eq:ELRdecomp} applies, and
compared the exact periodic $E_{\rm LR}$ with the open-boundary value of Eq.~\eqref{eq:open}. The
difference, between about $-0.6$ and $-1.0$~eV at $L=24$~\AA{} and about $-0.17$~eV at $L=120$~\AA, is
dominated by the Wigner term $C\xi_{\rm sc}Q^2/2L$ and reproduced by the leading terms of
Eq.~\eqref{eq:ELRdecomp} [Fig.~\ref{fig:identities}(e)]. The remainder falls from about $0.1$~eV at
$L=24$~\AA{} to about $2\times10^{-5}$~eV at $L=120$~\AA{} as $L^{-5}$ [Fig.~\ref{fig:identities}(f)], the
scaling of the leading anisotropic term ($\ell=4$) of a cubic cell; the visible deviations of the
Na$_8$Cl$_8^+$ clusters from the leading terms at $L\le32$~\AA{} are this remainder.

The LAMMPS implementation was compared with the TensorPotential reference, in which forces and virial
follow by automatic differentiation, for \pLES{} models of the two water datasets of Sec.~\ref{sec:water}:
for three cubic 192-atom cells of liquid water and a rhombohedral 36-atom ice~II cell ($a=7.78$~\AA,
$\alpha=\beta=\gamma=113.1^\circ$), energies agree to $7\times10^{-12}$~eV, forces to
$1.4\times10^{-9}$~eV/\AA{} (for forces up to 19~eV/\AA) and the virial to $4.3\times10^{-9}$~eV, with
identical results on one and four MPI ranks. In NVE molecular dynamics of the \pLES{} liquid-water model of
Table~\ref{tab:errors} (512 molecules, time step 0.5~fs, 100~ps), the total energy fluctuates by
$0.009$~meV/atom, about 75 times less than the kinetic energy, with no measurable drift
($|dE/dt|\approx0.01$~meV/atom/ns; Supplementary Material).

\section{Charged NaCl clusters}\label{sec:nacl}
\subsection{Data and models}
The NaCl cluster data of Ko \emph{et al.}~\cite{ko2021} were designed to expose non-local charge
redistribution. Na$_8$Cl$_8^+$ is obtained from Na$_9$Cl$_8^+$ by removing a neutral Na atom while the
remaining atoms are kept in place, so that both clusters carry a charge of $+1$ and the charge redistributes
over the whole cluster; a short-range potential cannot distinguish the resulting changes of the
potential-energy surface far from the removed atom~\cite{ko2021}. The reference data are spin-polarized PBE
calculations of isolated clusters with FHI-aims, together with Hirshfeld
charges~\cite{ko2021,perdew1996,blum2009,hirshfeld1977}.

I trained three models with identical short-range settings ($n_E=2$ FS densities, 150 basis functions per
element, cutoff 5~\AA) on 2000 structures drawn at random from the training data, using energies and
forces only, and evaluated them on 250 separate test structures: (i) ACE; (ii) ACE with an unconstrained
latent-charge channel, LES at $\mu=0$ [level (iv) of Proposition~\ref{prop:hierarchy}]; and (iii)
\pLES{}, the same model with the projection Eq.~\eqref{eq:projection}, $w_i=1$ and $Q_0=+1$ [level (iii)].
Both latent-charge models use $n_q=2$ charge densities and $\sigma=1$~\AA{} and evaluate $E_{\rm LR}$ with
Eq.~\eqref{eq:open}; for an isolated cluster in vacuum the gauge condition gives $\lambda=1$ and $Q_0=Z$.
LES ($\mu=0$) denotes the unconstrained formulation of Ref.~\cite{cheng2025les} implemented in ACE with the
same descriptor and training protocol as the other two models; it differs from the reference CACE-LES
implementation in the descriptor, the charge head and the training procedure. The Hirshfeld charges are
used only for comparison.

\subsection{Accuracy}
\begin{table}
\caption{Mean absolute errors / root-mean-square errors on the training and test sets, energies $E$ in
meV/atom and forces $F$ in meV/\AA. In each row, the lowest mean absolute error and the lowest
root-mean-square error are in bold. NaCl clusters: 2000 training and 250 test structures. Water
revPBE0-D3: 1291 training and 143 validation structures, the latter used for early stopping (shown as test).
Water revPBE-D3: 2386 training, 265 validation (early stopping) and 291 test structures.}
\label{tab:errors}
\footnotesize
\setlength{\tabcolsep}{3pt}
\begin{tabular}{llccc}
\toprule
Dataset & & ACE & LES ($\mu=0$) & \pLES{}\\
\midrule
\multicolumn{5}{l}{\emph{Training}}\\
Na$_9$Cl$_8^+$   & $E$ & 1.50 / 1.86 & 0.68 / 0.80 & \textbf{0.34} / \textbf{0.42}\\
                 & $F$ & 41.4 / 64.1 & 16.1 / 23.1 & \textbf{14.8} / \textbf{21.5}\\
Na$_8$Cl$_8^+$   & $E$ & 1.31 / 1.59 & \textbf{0.45} / 0.57 & 0.45 / \textbf{0.52}\\
                 & $F$ & 32.9 / 48.6 & 12.3 / 17.7 & \textbf{11.4} / \textbf{16.1}\\
NaCl, all        & $E$ & 1.41 / 1.73 & 0.56 / 0.69 & \textbf{0.39} / \textbf{0.47}\\
                 & $F$ & 37.2 / 57.2 & 14.3 / 20.7 & \textbf{13.2} / \textbf{19.1}\\
Water            & $E$ & 5.18 / 6.66 & 2.83 / 3.85 & \textbf{2.68} / \textbf{3.72}\\
revPBE0-D3       & $F$ & 67.7 / 112  & 55.6 / 90.6 & \textbf{52.0} / \textbf{85.5}\\
Water            & $E$ & 5.15 / 9.62 & 4.08 / 7.95 & \textbf{3.43} / \textbf{6.38}\\
revPBE-D3        & $F$ & 58.7 / 113  & 49.8 / 97.7 & \textbf{45.4} / \textbf{87.8}\\
\midrule
\multicolumn{5}{l}{\emph{Test}}\\
Na$_9$Cl$_8^+$   & $E$ & 1.52 / 1.91 & 0.73 / 0.84 & \textbf{0.37} / \textbf{0.45}\\
                 & $F$ & 41.2 / 64.3 & 15.9 / 23.2 & \textbf{14.7} / \textbf{21.4}\\
Na$_8$Cl$_8^+$   & $E$ & 1.35 / 1.63 & \textbf{0.42} / \textbf{0.54} & 0.47 / 0.55\\
                 & $F$ & 32.8 / 48.4 & 12.4 / 17.9 & \textbf{11.5} / \textbf{16.3}\\
NaCl, all        & $E$ & 1.44 / 1.77 & 0.58 / 0.71 & \textbf{0.42} / \textbf{0.51}\\
                 & $F$ & 37.2 / 57.2 & 14.2 / 20.8 & \textbf{13.2} / \textbf{19.2}\\
Water            & $E$ & 6.11 / 7.63 & 3.26 / 4.24 & \textbf{2.98} / \textbf{4.01}\\
revPBE0-D3       & $F$ & 70.6 / 128  & 58.3 / 98.7 & \textbf{54.9} / \textbf{94.7}\\
Water            & $E$ & 5.67 / 11.1 & 5.12 / 12.0 & \textbf{4.15} / \textbf{9.30}\\
revPBE-D3        & $F$ & 51.8 / 114  & 47.6 / 181  & \textbf{41.4} / \textbf{107}\\
\bottomrule
\end{tabular}
\end{table}

Training and test errors are close for all models, except for the root-mean-square force errors on the
water phases (Sec.~\ref{sec:water}).
Table~\ref{tab:errors} gives the training and test errors (parity plots in the Supplementary Material).
The latent-charge channel reduces the mean absolute energy error of ACE by a factor of 2.5 without and 3.4
with the projection, and the force error by a factor of 2.6 and 2.8, respectively. The projection lowers
the energy error of Na$_9$Cl$_8^+$ from 0.73 to 0.37~meV/atom and leaves that of Na$_8$Cl$_8^+$ nearly
unchanged. For orientation, Ref.~\cite{rinaldi2025} reports root-mean-square errors of 1.55~meV/atom and
41.7~meV/\AA{} for ACE and 0.57--0.71~meV/atom and 12--18~meV/\AA{} for charge-equilibrated ACE on these
data, with a different training set and protocol and with Hirshfeld charges in the training; the
corresponding root-mean-square errors of the present models are 1.77, 0.71 and 0.51~meV/atom and 57, 21
and 19~meV/\AA{} for ACE, LES ($\mu=0$) and \pLES{}.

\subsection{Latent charges}
\begin{figure*}
  \includegraphics[width=\textwidth]{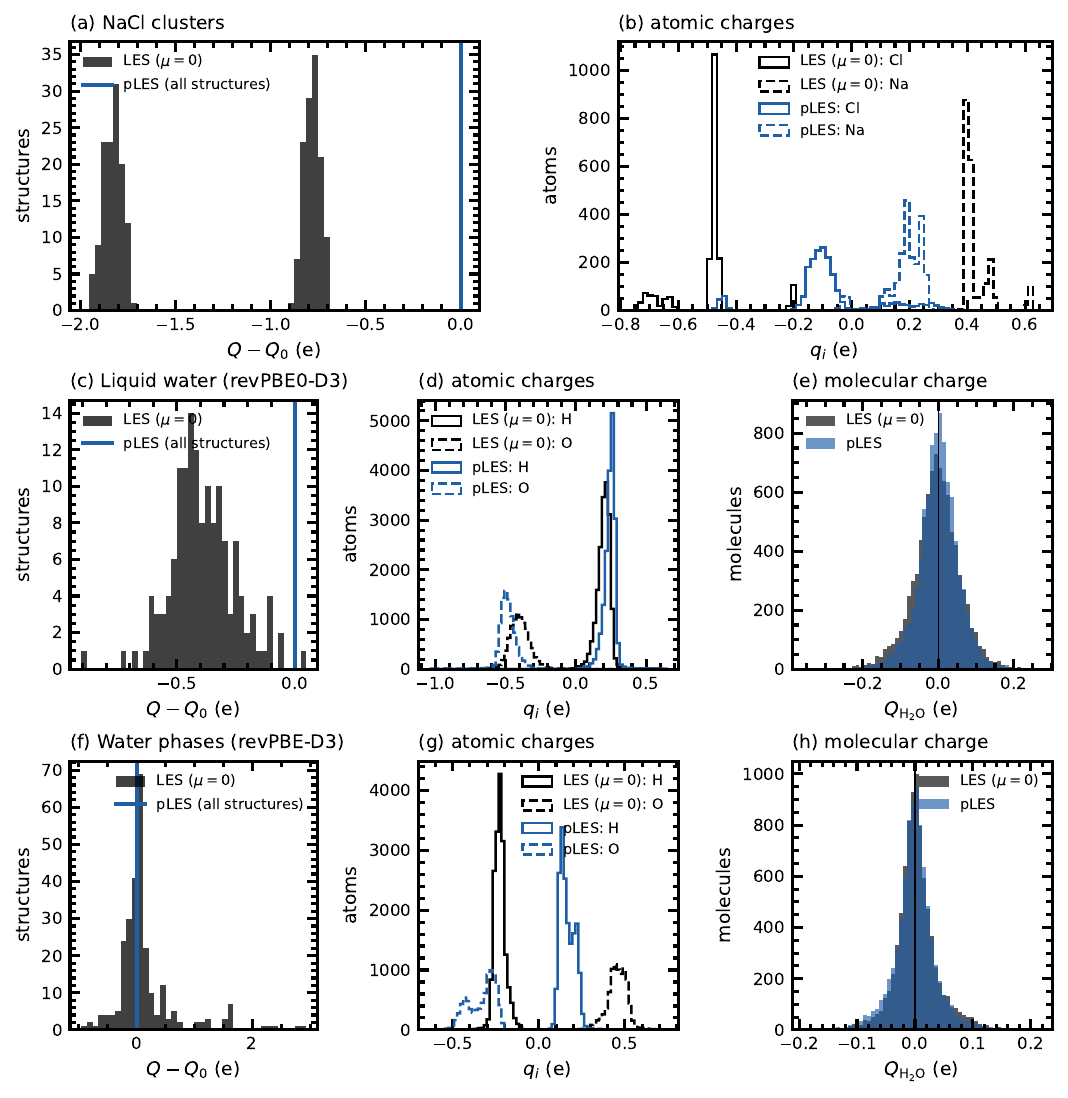}
  \caption{Latent charges of LES ($\mu=0$) and \pLES{} on the test structures of the NaCl clusters (a, b)
  and the validation structures of liquid water (revPBE0-D3; c--e) and of the water phases (revPBE-D3),
  including ices, high-pressure phases and diatomic curves (f--h). (a, c, f) Deviation of the total latent
  charge from $Q_0$ per structure: \pLES{} satisfies $Q=Q_0$ exactly (vertical line), LES ($\mu=0$) does
  not. (b, d, g) Atomic latent charges by species; the overall sign of the LES ($\mu=0$) charges is
  arbitrary (compare d and g). (e, h) Latent charge of intact water molecules.}
  \label{fig:latent_charges}
\end{figure*}

\paragraph*{Total charge.} Without the projection, the latent charges of each cluster sum to
$Q-Q_0\approx-0.8\,e$ for Na$_9$Cl$_8^+$ and $\approx-1.8\,e$ for Na$_8$Cl$_8^+$
[Fig.~\ref{fig:latent_charges}(a)]: the unconstrained model removes about one unit of latent charge together
with the Na atom, as any sum of local terms must, although the physical charge is $+1$ in both clusters.
Its accuracy is nonetheless close to that of \pLES{}, so energies and forces do not determine the net charge.
In a periodic cell this net charge would produce the artefact $\frac C2(\psi_\Lambda+2\pi\sigma^2/V)Q^2$ of
Eq.~\eqref{eq:ELRdecomp} and, because $Q$ fluctuates, spurious forces. With the projection, $Q=+1$ to
machine precision for every structure. Within the hierarchy of Sec.~\ref{sec:qeq}, the single chemical
potential of the diagonal limit supplies the non-local information that this example requires, at the
cost of one global sum and without a linear solve.

\paragraph*{Partition and dipoles.} The \pLES{} charges correlate only partially with the Hirshfeld charges
($q_i=s\,q_i^{\rm H}$ with $s=0.32$, $R^2=0.40$, Pearson $r=0.68$), and some Cl atoms carry a positive latent
charge; the redistributions $q_i-Q/N$ of LES ($\mu=0$) and \pLES{} agree only partially as well ($r=0.61$,
Fig.~\ref{fig:compare}). The cluster dipoles of \pLES{} about the centroid are linearly related to the
Hirshfeld dipoles but with the opposite orientation ($r=-0.98$; Supplementary Material). Reversing the
redistribution, $q_i-Q_0/N\to-(q_i-Q_0/N)$, changes $E_{\rm LR}$ by 16--26~meV/atom on average, far more
than the model error, so the orientation is fixed jointly with $E_{\rm loc}$ rather than by the reference
charges. Within a charged cluster, energies and forces thus fix the net charge, which the projection
supplies, but neither the atomic partition nor the intra-cluster redistribution, in line with
Sec.~\ref{sec:gauge}.

\section{Water}\label{sec:water}
\subsection{Data and models}
Two datasets are used. The liquid-water data of Ref.~\cite{cheng2019water} contain 1593 configurations of
64 molecules computed with revPBE0-D3; I split them into 1291 training, 143 validation and 159 test
configurations and used the validation set for early stopping.
The water-phase data of
Ref.~\cite{ibrahim2024efficient, ibrahim2026pd} contain liquid water, ices, high-pressure phases, active-learning
configurations and diatomic H$_2$, OH and O$_2$ curves computed with revPBE-D3; 10\% of the training
structures of each group were set aside for validation, giving 2386 training, 265 validation and 291 test
structures.
The three models of Sec.~\ref{sec:nacl} were trained with cutoffs of 5.5~\AA{} (liquid water) and
5.0~\AA{} (water phases), $\sigma=1$~\AA{}, $k_c=\pi$~\AA$^{-1}$ and $Q_0=0$, with early stopping on the
validation set for both datasets.

\subsection{Accuracy and latent charges}
For liquid water, both latent-charge models reduce the errors of ACE, from 6.1 to 3.0--3.3~meV/atom and
from 71 to 55--58~meV/\AA{} (Table~\ref{tab:errors}). For the water phases, \pLES{} has the lowest errors in
all four measures, reducing those of ACE from 5.67 to 4.15~meV/atom and from 51.8 to 41.4~meV/\AA{} in mean
absolute error and from 11.1 to 9.30~meV/atom and from 114 to 107~meV/\AA{} in root-mean-square error.
LES ($\mu=0$) reduces the mean absolute errors as well, but its root-mean-square force error, 181~meV/\AA,
exceeds that of ACE; its ratio to the mean absolute error shows that it is dominated by a few structures
with large force errors. In neutral systems the purpose of the projection is physical consistency, the
correct net charge and no periodic artefact; the lower errors of \pLES{} on the water phases are an
additional, empirical observation.

\begin{figure}
  \includegraphics[width=\columnwidth]{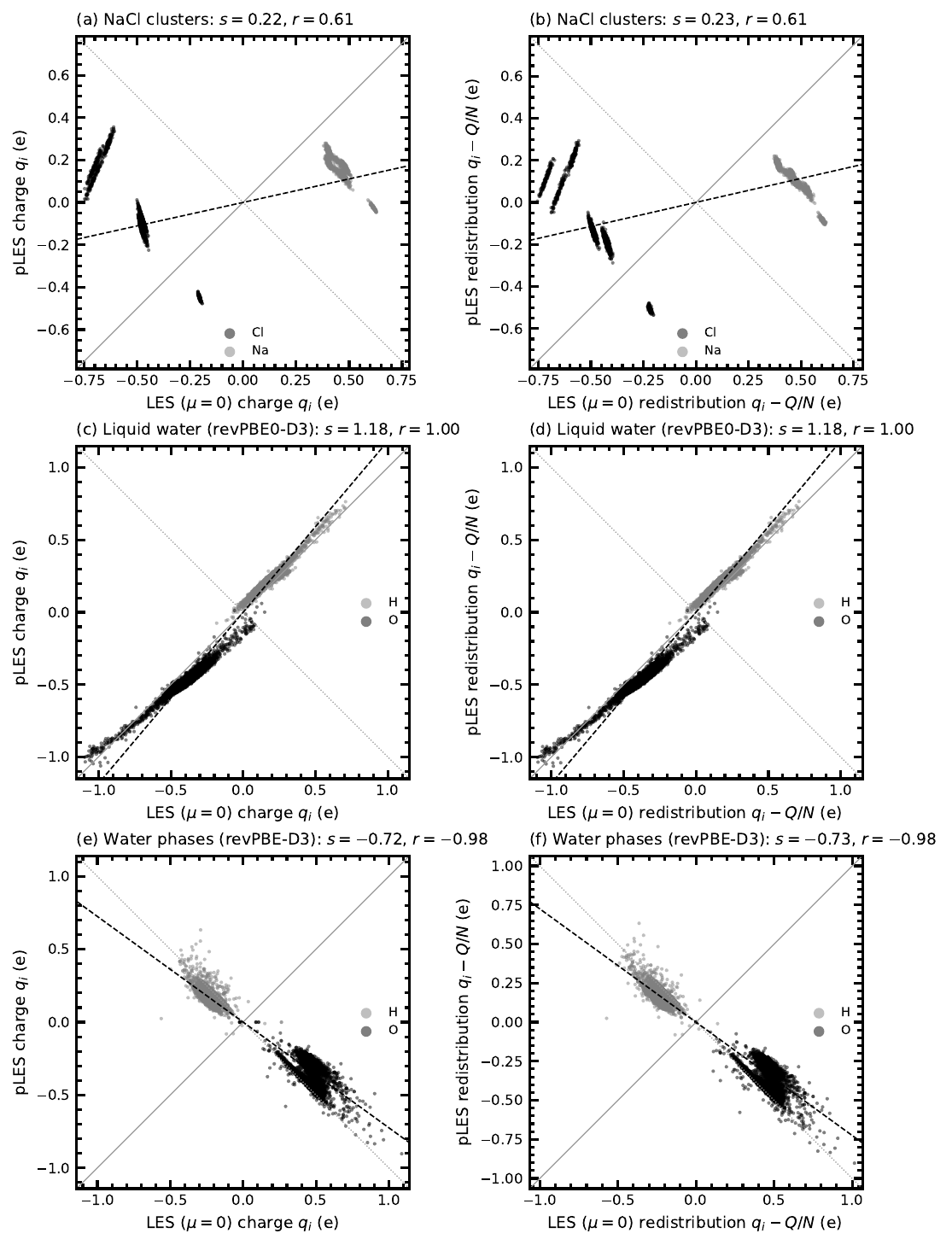}
  \caption{Latent charges of \pLES{} against LES ($\mu=0$) on the same structures (test set for the NaCl clusters, validation sets for water), atom by atom: raw
  charges (left) and redistributions $q_i-Q/N$ (right) for the NaCl clusters, liquid water and the water
  phases. Dashed: least-squares slope $s$ through the origin; thin lines: $\pm$ diagonal.}
  \label{fig:compare}
\end{figure}

The unconstrained model places a net charge on every neutral cell [Fig.~\ref{fig:latent_charges}(c, f)]:
on average $-0.39\,e$ per 192-atom cell of liquid water and $+0.14\,e$ per structure in the water
phases, varying from structure to structure. \pLES{} keeps
every cell exactly neutral, and the molecular charges of both models are centred on zero
[Fig.~\ref{fig:latent_charges}(e, h)]. Figure~\ref{fig:compare} compares the two models directly. In both
water datasets they learn the same latent charges up to a global sign and scale: $r=1.00$ with $s=1.18$ for
liquid water and $r=-0.98$ with $s=-0.72$ for the water phases, where the unconstrained model has negative
hydrogen charges. This is the freedom of Sec.~\ref{sec:gauge}: the sign of the unconstrained model is set by
its initialization, and the scale is set jointly with $E_{\rm loc}$. Apart from it, the projection removes
exactly the part of the charges that energies and forces cannot fix and that a periodic cell penalizes,
the net charge, and leaves the learned charge pattern unchanged. In the charged clusters, by contrast, the
redistribution itself is not determined ($r=0.61$).

\subsection{Long-wavelength response of the latent channel}
I ran $NVT$ molecular dynamics of the \pLES{} liquid-water model of Table~\ref{tab:errors} with the LAMMPS
implementation at 300~K and a time step of 0.5~fs in cubic cells of 512 molecules ($L=24.87$~\AA, 4.0~ns)
and 1728 molecules ($L=37.30$~\AA, 1.0~ns), with positions and latent charges stored every 100~fs; the first
20~ps were discarded and every second frame was analysed. 
The total latent charge stays zero to the precision of the stored charges ($10^{-4}\,e$). The
latent charges of the molecules  average to zero but fluctuate by $0.039\,e$ in both cells:
\pLES{} water has no exact local charge conservation. As a control with locally conserved charge
[case (a) of Proposition~\ref{prop:screening}], I ran rigid SPC/E water~\cite{berendsen1987} (SHAKE
constraints, Lennard-Jones and real-space Coulomb cutoff 10~\AA, particle--particle particle--mesh
electrostatics with relative accuracy $10^{-5}$ and tin-foil boundaries) in the same two cells at 300~K,
with a Nos\'e--Hoover thermostat and a time step of 2~fs, for 5~ns in both cells after 110~ps of equilibration, and analysed frames every 200~fs with the point-charge kernel
$v(k)=4\pi C/k^2$.

\begin{figure}
  \includegraphics[width=\columnwidth]{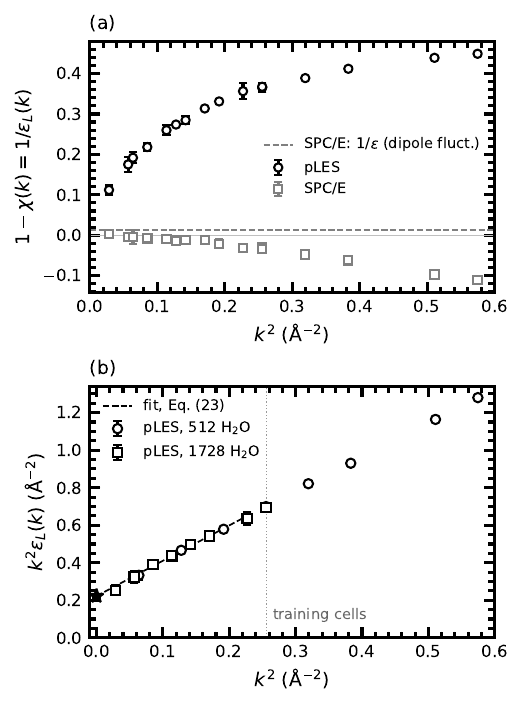}
  \caption{Long-wavelength charge response of \pLES{} water (latent charges) and of SPC/E water (fixed
  charges) from MD in cells of 512 and 1728 molecules. (a) $1-\chi(k)=1/\varepsilon_L(k)$,
  Eq.~\eqref{eq:sumrule}; dashed: $1/\varepsilon$ of SPC/E from its dipole fluctuations. (b)
  $k^2\varepsilon_L(k)$ of \pLES{} against $k^2$, with the fit of Eq.~\eqref{eq:conductor} to the lowest ten
  $k^2$ (dashed) and its intercept $k_s^2$ (star); a dielectric whose $\varepsilon_L(k)$ has converged would
  extrapolate to the origin.
  Vertical dotted line: smallest $k^2$ of the 64-molecule training cells. Error bars: 95\% intervals over
  five blocks.}
  \label{fig:eps}
\end{figure}

The two cells give one curve where their wavevectors are close, so $\chi(k)$ is a property of the liquid
rather than of the cell. $\chi$ rises monotonically towards $k=0$, to $0.888\pm0.011$ at the smallest
wavevector, $k=0.168$~\AA$^{-1}$; correspondingly $\varepsilon_L(k)=1/(1-\chi)$ grows from about 2 at
$k=0.76$~\AA$^{-1}$ to 8.9 [Fig.~\ref{fig:eps}(a)], and $k^2\varepsilon_L(k)$ falls almost linearly with
$k^2$ towards a positive intercept [Fig.~\ref{fig:eps}(b)]. Fitted to the lowest ten $k^2$ of both cells
($k\le0.48$~\AA$^{-1}$, 8 degrees of freedom), a dielectric whose $\varepsilon_L(k)$ has converged at these
wavevectors, $\chi=1-\varepsilon_{\rm lat}^{-1}+ak^2$, is excluded ($\chi^2=133$).
Equation~\eqref{eq:conductor} gives $\varepsilon_d=1.9$ and $k_s^2=0.222\pm0.011$~\AA$^{-2}$ ($\chi^2=13.4$).
The Debye estimate for uncorrelated molecular charges, $4\pi\beta Cn\langle Q_A^2\rangle=0.36$~\AA$^{-2}$, is
of the same order; the smaller fitted value is consistent with partial anticorrelation of the charges of
neighbouring molecules.

The $k$ dependence nevertheless resolves a crossover. Fitting
$1-\chi=\varepsilon_{\rm lat}^{-1}+k^2/(k_s^2+\varepsilon_dk^2)$ to the same points gives $\chi^2=3.4$ and
$\varepsilon_{\rm lat}^{-1}=0.034$ (95\% interval 0.014--0.052), i.e.\ $\varepsilon_{\rm lat}\approx29$ with
$19\le\varepsilon_{\rm lat}\le71$. The conductor limit $\varepsilon_{\rm lat}^{-1}=0$, Eq.~\eqref{eq:conductor},
is excluded ($\Delta\chi^2=10.1$, $p\approx0.002$); with ten blocks instead of five,
$\varepsilon_{\rm lat}^{-1}=0.014$--$0.061$ and $\Delta\chi^2=8.4$. The crossover wavevector
$k_s/\sqrt{\varepsilon_{\rm lat}}\approx0.09$~\AA$^{-1}$ lies below the smallest sampled wavevector, so
$\varepsilon_{\rm lat}$ is an extrapolation within the assumed form and rests mainly on the smallest
wavevectors of the larger cell. In real space, the molecular-charge fluctuations are compensated over
$\sqrt{\varepsilon_{\rm lat}}/k_s\approx1.2$~nm (0.9--1.8~nm), several molecular shells rather than nearest
neighbours, for which the $1/k^2$ regime would end near $k\approx0.3$~\AA$^{-1}$. The 64-molecule training
cells ($L\approx12.4$~\AA) contain no wavevector below $k\approx0.5$~\AA$^{-1}$, so their energies and forces
cannot constrain this regime: the finite $\varepsilon_{\rm lat}$ is a property of this trained model, not a
consequence of the charge model [Proposition~\ref{prop:screening}(b)].

The SPC/E control calibrates this analysis [Fig.~\ref{fig:eps}(a)]. Its molecules are exactly neutral, and
the dipole fluctuations give $\varepsilon=77\pm9$ and $72\pm6$ in the two cells. At the smallest wavevectors
$\chi=1.004\pm0.018$ and $0.997\pm0.007$, consistent with $1-\varepsilon^{-1}\approx0.986$: the two routes
agree, as Proposition~\ref{prop:screening}(a) requires. Because the response of SPC/E has converged at these
wavevectors, $\chi(k)$ alone bounds its large $\varepsilon$ only from below; for \pLES{} the bounds come
instead from the $k$ dependence of $\chi$ and depend on the crossover form. The shape of the response
differs qualitatively. SPC/E has $1-\chi\le0$ within error at all sampled wavevectors and $\chi$
significantly above 1 for $k\gtrsim0.3$~\AA$^{-1}$, the overscreening of a polar liquid. The latent channel
of \pLES{} has $1-\chi$ between 0.11 and about 0.45, positive and decreasing monotonically towards zero; at
$k=0.168$~\AA$^{-1}$ it screens far more weakly than SPC/E ($\varepsilon_L=8.9$ against
$\varepsilon_L\gtrsim100$).

Neutralizing each molecule removes the molecular charges but not their correlations with the dipoles. The
dipole fluctuations of the neutralized molecules give $\varepsilon=19.7\pm0.8$ and $17.6\pm2.4$ in the two
cells (Supplementary Material). As stated in Proposition~\ref{prop:screening}, this is not a dielectric
constant of the model: the fluctuating molecular charges, anticorrelated with the molecular dipoles, are
part of the response, and its similarity to the fitted $\varepsilon_{\rm lat}$ is not required by the theory.

Because $\varepsilon$ of water is large, these differences matter for long-range quantities rather than for
short-range structure: the gauge relation Eq.~\eqref{eq:gauge} and finite-size corrections that assume a
finite $\varepsilon$ require a finite $\varepsilon_{\rm lat}$. Here the data give one, but only within an
assumed crossover form and for one trained model. A finite $\varepsilon_{\rm lat}$ can be guaranteed only
through local charge conservation in the charge model, case (a) of Proposition~\ref{prop:screening}: a
projection onto fixed charge for each molecule rather than for the whole cell, or charges written as
antisymmetric transfers $q_i=\sum_jp_{ij}$, $p_{ij}=-p_{ji}$, between neighbouring atoms, as in split-charge
equilibration~\cite{nistor2006}. The second form also applies to reactive systems; both are left for future
work.

\section{Conclusions}\label{sec:conclusions}
Writing the long-range energy of latent-charge potentials as a smeared Epstein--Hurwitz sum makes its
periodic structure exact: the net-charge artefact, the Gaussian self-energy, the sphere-averaged pair
potential, the forces and the virial follow from one decomposition, which holds to machine precision and
for the charges of trained models. The same formulation places the charges of LES and of charge
equilibration in one hierarchy, in which a projection of the latent charges onto fixed total charge is the
diagonal-hardness limit and costs one global reduction. \pLES{} combines these elements and is implemented
in ACE and LAMMPS.

Energies and forces determine the latent charges only up to a gauge. The overall sign is set by the
initialization in neutral systems, the scale and the atomic partition jointly with the local energy, and the
net charge not at all without the projection. Independently trained models with and without the projection
learn the same charges in water up to sign and scale, while the unconstrained model misses the net charge
of charged clusters by up to 1.8~$e$ and puts a spurious net charge on neutral cells. Fragment-level
quantization, $Q_A^{\rm lat}=\lambda z_A$, is therefore the consistent target for latent point charges.

The long-wavelength response sharpens this statement. The latent channel is guaranteed to be a dielectric
only if charge is conserved locally; latent charges that are local functions of the environment and are
conserved at most globally, as in \pLES{}, unconstrained LES and charge equilibration with one chemical
potential, can screen like a conductor.
In molecular dynamics of \pLES{} water, the molecular latent
charges fluctuate and dominate the screening at all accessible wavelengths; within a crossover model, the
data exclude a conductor and give a dielectric constant of the latent channel between 19 and 71, set by
compensation of molecular charges over about 1~nm. Because training cells do not probe these wavelengths,
this is a property of the trained model; a finite dielectric constant can be guaranteed only by the charge
model, through fragment-wise projection or charge transfer between neighbours.
The consequences for potentials
of mean force, charge-changing reactions and finite-size corrections in periodic cells are the subject of
future work.

\section*{Supplementary Material}
Training protocol; energy and force parity plots for all datasets; latent vs Hirshfeld charges and cluster
dipoles for the NaCl clusters; energy conservation in $NVE$ molecular dynamics; details of the
long-wavelength analysis with the dielectric, conductor and crossover fits for five and ten blocks; the
response of the neutralized molecules; the SPC/E control.

\section*{Data Availability}
The NaCl cluster data are available from Ref.~\cite{ko2021}, the liquid-water data from
Ref.~\cite{cheng2019water} and the water-phase data from Ref.~\cite{ibrahim2024efficient, ibrahim2026pd}.
The fitted
models, training inputs and analysis scripts are available upon acceptance.

\section*{Code Availability}
The implementation in TensorPotential/\textsc{pacemaker} and the LAMMPS pair style \texttt{pace/les} will be
released as open source with the companion software publication.

\begin{acknowledgments}
I acknowledge institutional support from KAUST. For computer time, this research used Shaheen III and Ibex managed by the Supercomputing Core Laboratory at King Abdullah University of Science and Technology (KAUST) in Thuwal, Saudi Arabia.
\end{acknowledgments}

\bibliography{refs}

\end{document}